# Interpreting Recoil For Undergraduate Students

Tarek A. Elsayed

*Electronics Research Institute, Dokki, Giza 12622, Egypt*
*and*
*Institute of Theoretical Physics, University of Heidelberg, Philosophenweg 19, 69120 Heidelberg, Germany*

tarek.ahmed.elsayed@gmail.com

The phenomenon of recoil is usually explained to students in the context of Newton's third law. Typically, when a projectile is fired, the recoil of the launch mechanism is interpreted as a reaction to the ejection of the smaller projectile, since "Each action has an equal and opposite reaction". The same phenomenon is also interpreted in the context of the conservation of linear momentum, which is closely related to Newton's third law. Since the actual microscopic causes of recoil differ from one problem to another, some students (and teachers) may not be satisfied with understanding recoil through the principles of conservation of linear momentum and Newton's third law. For these students, the origin of the recoil motion should be presented in more depth.

A survey conducted at King Fahd University of Petroleum and Minerals in the form of a quiz taken by more than 120 freshman students about the cause of recoil motion in different systems revealed that the justification of recoil as given by many students is no more than "an action and a reaction" without a real understanding of how the reaction is produced in different, specific problems.

The most famous examples of the recoil phenomenon are rocket propulsion or the firing of a bullet from a rifle. In a review that included a dozen textbooks of introductory physics, I discovered that the explanation of recoil in all of them goes no deeper than describing that the rocket (rifle) exerts a strong force on the gas (bullet), expelling it, while the gas (bullet) exerts an equal and opposite force on the rocket (rifle) in accordance with the conservation of momentum [1]. Therefore, many students are unaware that the compressed gas expanding in all directions inside the rocket actually produces the force that propels the rocket rather than the projectile itself. Questions such as how and when these recoil forces appear and where they originate are not easy to answer using the descriptive interpretation of recoil as presented in many textbooks. This is even more evident in complex systems such as the rail gun where the origin and distribution of the recoil forces have been a research topic of the last two decades.[2] Allowing students to think more deeply about the origin of such a simple phenomenon, down to the most microscopic level, will lead them to seek the same level of deep understanding of other physical systems.

To explain recoil in firing systems to students, the instructor can use a model consisting of a movable container filled with pressurized gas and capped with a movable piston as shown in Fig. 1. Examples of other systems are mentioned at the end of this article. The piston is fixed to the container with bolts. When these bolts are released, the piston will be pushed to the left while the container recoils to the right in order to conserve the total linear momentum.

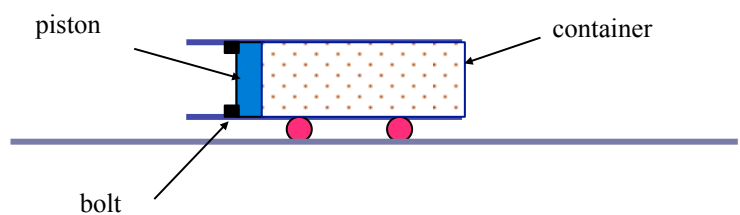

Figure 1

The descriptive explanation of the recoil, based solely on Newton's third law or the conservation of linear momentum, may lead some students to inquire how or when the container "knows" about the ejection of the piston. As was revealed by the answers to the quiz, 17% of students even think that some delay exists between the motion of the piston and container. Another issue that proves the limitation of the descriptive interpretation to answer all the students' questions is the proper geometric relation between a "large object" and a "small object" for recoil to occur when the latter is fired from the former. Should the "large object" merely enclose the "smaller" one? Or more generally: is there a general and more precise definition of the connection between the two objects that dictates when the recoil should occur and when it should not? For example, consider the case of a man firing a gun. The gun itself recoils and the man recoils only after being pushed by the gun. Students may ask: "Why doesn't the man recoil earlier in the same instant that the gun starts its backward motion? Isn't he enclosing the bullet as well?" To test the students' understanding of this issue, a small complication was added to the example in Fig.1 by enclosing the gas container in a large box that can move freely in a horizontal direction as shown in Fig. 2. The students were asked whether the large box would recoil relative to a stationary frame when the piston was released. The argument of the conservation of linear momentum alone is insufficient to answer this question, since we have a single equation and two degrees of freedom: the speed of the gas container and the speed of the outer box. Similarly, the argument that recoil occurs merely as a reaction to the ejection of the piston as dictated by Newton's third law does not explain why the outer box cannot participate in this reaction. Therefore, it is not surprising that approximately 40% of the students answered this question affirmatively.

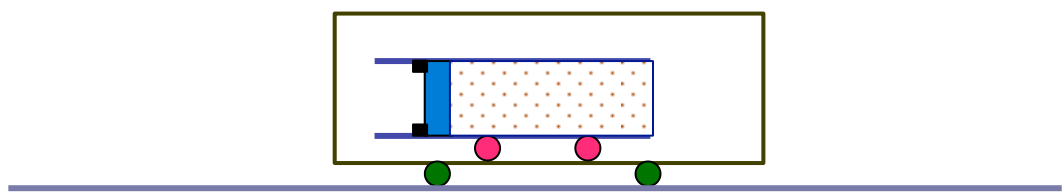

Figure 2

An effective way to explain recoil in such a system is to get the students to realize that the gas was exerting equal forces on the opposite sides of the container in Fig. 1 owing to the bombardment between the gas molecules and the walls before the release of the bolts. Once the bolts are released, the force on the piston will push it to the left while the uncompensated force on the other side of the container will push it to the right. More simply, when the bolts are released, the gas tends to expand and therefore pushes the piston and container away from each other. Being introduced to the concept of recoil in this way, students realize that recoil originates from the rear of the container and not from the side walls. The launching of the piston and the recoil of the container occur approximately at the same instant. The only delay that is involved here is the propagation time of the alteration of the electrostatic force between the neighboring atoms of the sidewalls of the container. This alteration will be transferred from the contact points with the bolts to the rear of the container after the release of the bolts. Since electrostatic disturbances transfer from one point to another by the speed of light, this particular time is of the order of the length of the container divided by the speed of light, which is negligible. From this perspective, a student may consider the motions of the piston and container as recoils to one another.

Once the actual cause of recoil is explained to students, they understand that only the gas container, as shown in Fig. 2, will recoil as the gas is acting by a force on its walls, and no force is acting on the large box.

Two simple experiments can be conducted in the classroom in order to explain this concept to students. First, an instructor can use a tank filled with water that is capable of moving freely, as shown in Fig. 3.

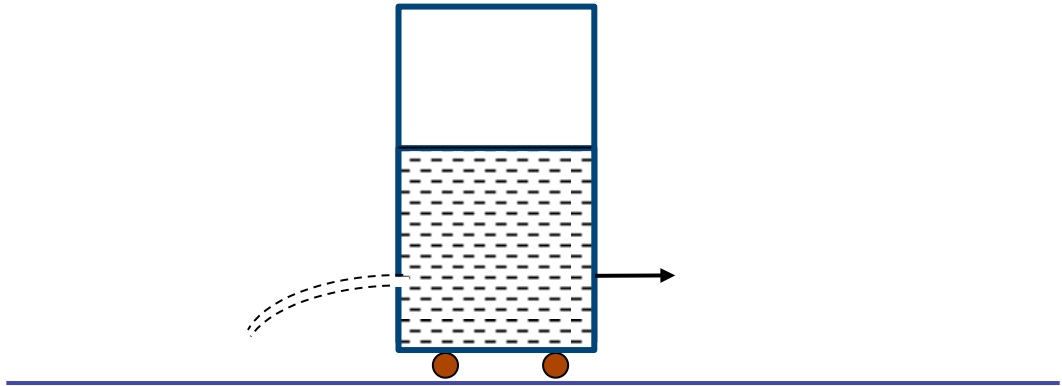

Figure 3

When a hole is made in one side of the tank, water will flow out of the tank and the tank will recoil in the other direction. Similarly to the gas container, water is acting by a force on all sides of the container because of its pressure. When a hole is made in the left wall, the pressure decreases at the hole and thereby the total force acting on the left wall becomes less than the right one. The recoil of the tank occurs because of these unbalanced forces on the tank. Another simple demonstration of recoil can be shown by using a fire extinguisher exhausting $CO_2$. In this demo, the instructor can let a student hold the fire extinguisher and ride on a cart or a bicycle as shown in Fig. 4.

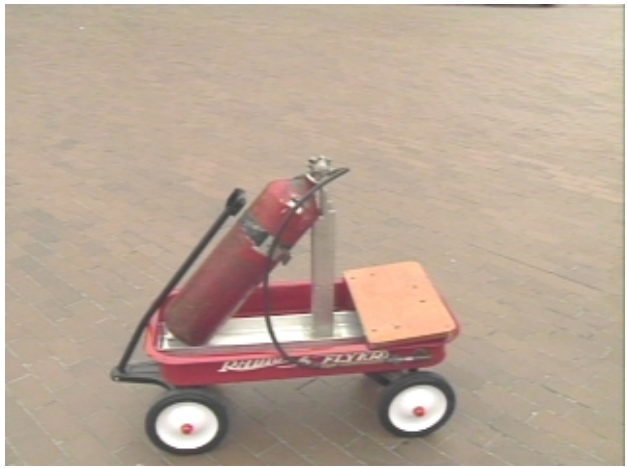 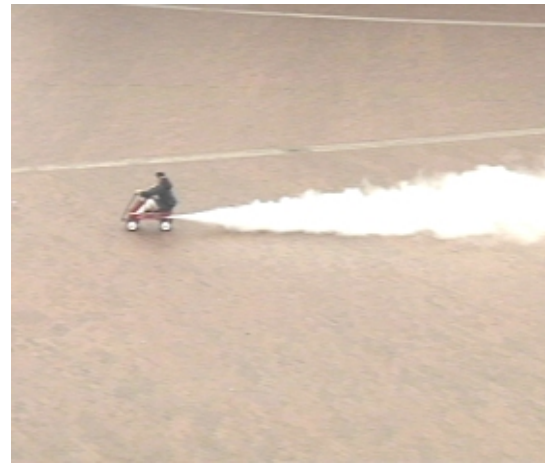

Figure 4
*(Courtesy of The Video Encyclopedia of Physics Demonstrations)*

When the student ignites the fire extinguisher, he or she will be propelled in the opposite direction to the flow of carbon dioxide, in turn demonstrating a classroom implementation of the rocket engine. Once the recoil in firing systems is fully explained to students, other systems undergoing recoil can be introduced to them in the form of deep, thought-provoking exercises [4]. Four such systems are introduced in the supplement attached to this article.

After the aforementioned quiz, the concept of recoil was introduced to 18 students in the way described in the present paper and most of the previous examples were discussed with them. The students were receptive to the discussion and almost all of them said that they learned something new and benefited from the discussion. Having learned to describe recoil at the deepest level, students should be able to apply the same line of thinking in other physical phenomena. Going so far in attributing causes to effects may foster a generation of physicists that unfolds new horizons of science and its applications.

**Acknowledgments**

The comments of Dr. Z. Yamani and Prof. M. A. Gondal from King Fahd University of Petroleum and Minerals and the discussion with Mati Meron from the University of Chicago are highly appreciated. The comments and suggestions of an anonymous referee are sincerely appreciated as well. The author would like to thank Dr. A. Alkarmi, Prof. M. A. Gondal, and Mr. A. Alaswad for conducting the quiz with their respective students.

# Supplement

In this supplement, extra examples are introduced to enhance the pedagogical method introduced above. The first of these systems is the rotary lawn sprinkler. According to the traditional way of explaining recoil, the rotation of the sprinkler, as shown in Fig. 5, could simply be attributed to the conservation of angular momentum or be justified as being a reaction to the flow of water from the sprinkler.

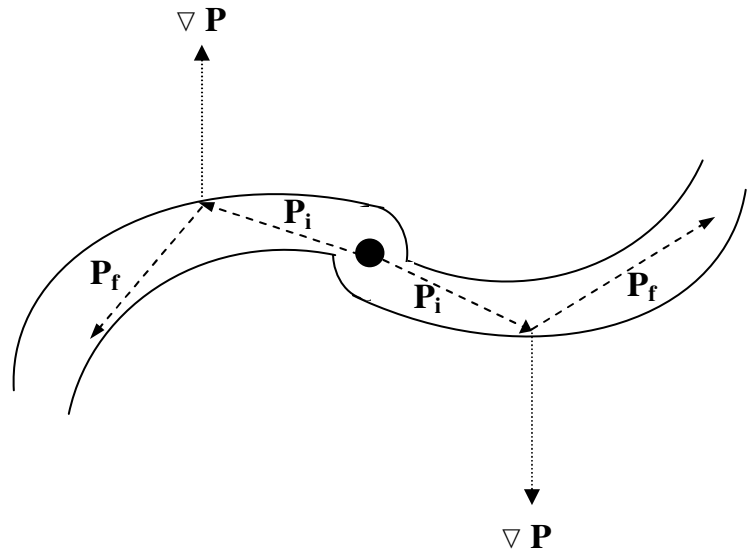

Figure 5.

The instructor can challenge his or her students to search for the actual force that causes the sprinkler to rotate opposite to the water flow or a hose to recoil when the tape connected to it is opened. The force in both cases is simply the push of water on the sidewall of the sprinkler or the hose. The flowing water in the right arm of the sprinkler pushes the lower side of the arm downward while the flowing water in the left arm pushes the upper side upward. These pushes cause the sprinkler to rotate clockwise. In taking a snapshot of the rotating sprinkler, the arrows in Fig. 5 indicate the directions of the water momentum before and after colliding with the walls as well as the direction of the change in water momentum ($\Delta P$). Making the arms of the sprinkler straight or the body of the hose fully stretched, one will find that none of them will recoil.

The second example is a swing connected to a stand that is able to move freely in a horizontal direction, as shown in Fig. 6. When the pendulum starts to oscillate, the stand will move back and forth out of phase with the oscillation of the swing in such a way as to keep the total linear momentum at zero. As before, the oscillation of the swing can be explained as owed to the conservation of momentum or justified as being a reaction to the swinging of the pendulum. The real force, however, that pushes the swing back and forth is the component of the gravitational force on the bob along the direction of the rod. The other component in the orthogonal direction simply rotates the pendulum and has no effect on the stand since the arm can move freely in the angular direction.

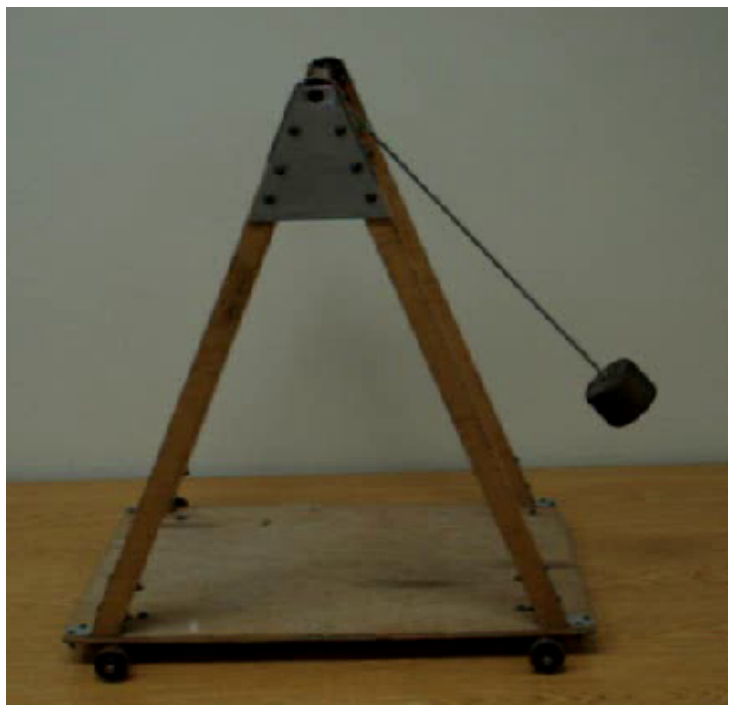

Figure 6

From the physics demonstration room, King Fahd University of Petroleum and Minerals

The third example is the recoil of a DC motor at its start-up. When current is switched on in a DC motor, the rotor starts to rotate and we can see that the stator starts to rotate momentarily in the reverse direction, stopped only by an outside effect, such as the force of friction. The instructor can demonstrate this process to his or her students by a simple DC motor from a motorized toy. As before, students should seek reasoning for recoil deeper than the descriptive argument of the conservation of angular momentum. The force that causes the stator to rotate is a result of the interaction between the magnetic field of the rotor and the magnetic dipoles in the stator in the case of a permanent magnet stator. To illustrate this, one can use a single current loop to represent the magnetic effect of each pole in the stator, as shown in Fig. 7-a.

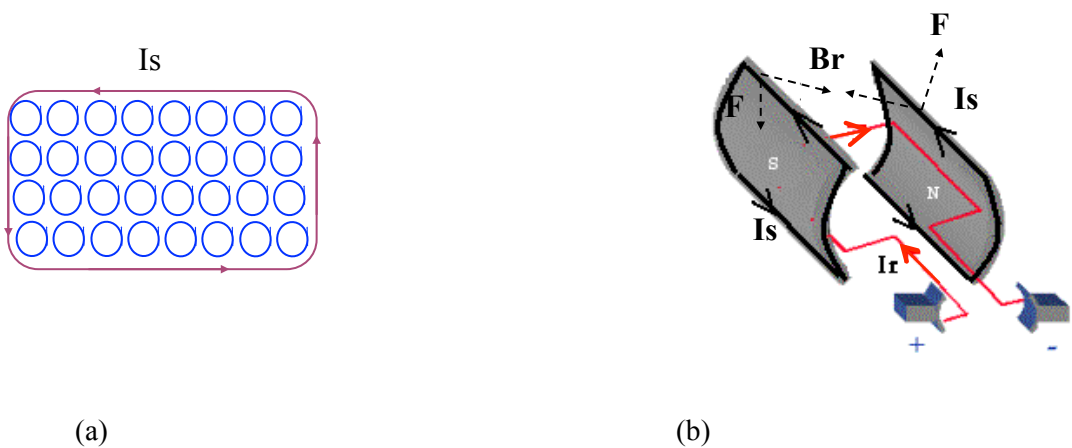

(a)  (b)

Figure 7.

The current of the rotor (**Ir**) produces a magnetic field (**B**) that acts by means of a force whose magnitude and direction are given by the Biot and Savart Law on the equivalent current of stator (**Is**). By assuming the rotor to be very long, we can only consider the interaction between the currents in the branches along the shaft of the rotor. In the configuration shown in Fig.7-b [3], when the current is switched on in the rotor, it will start to rotate clockwise. At the same time, the magnetic field generated by the rotor will act by a net force on the north pole current loop upwards and on the south pole current loop downwards in turn, causing the stator to rotate counterclockwise. This leads to the conservation of the angular momentum of the whole motor. The last example is the recoil of an atom when a photon is spontaneously emitted. The recoil can simply be attributed to the disturbance in the electrostatic force between the nucleus and electron, as the mean distance between them changes during transition. Since the electron undergoes a change of its wave function, the mean distance between the electron and nucleus is different before and after the transition and therefore the force affecting the nucleus is perturbed.